\documentstyle[twoside,fleqn,espcrc2,epsfig,graphics]{article}

\newcommand{\AmS}{{\protect\the\textfont2
   A\kern-.1667em\lower.5ex\hbox{M}\kern-.125emS}}

\title{Fragmentation functions using e$^+$e$-$ data from PETRA and LEP}

\author{ M. Blumenstengel $^{\mathrm{a}}$ and  S. Kluth 
\address{Max Planck Institute of Physics (Werner Heisenberg Institute), \\
			F\"ohringer Ring 6, 80805 M\"unchen, Germany }}

\begin{document}

\begin{abstract}

We investigate fragmentation of charged
particles in e$^+$e$^-$ annihilation at 22, 35, and 44 GeV
in terms of their polar angle and momentum distributions. 
From the angular distribution the ratio of longitudinal 
to total hadronic cross-section was determined 
at an average energy scale of 36.6 GeV to be $\sigma_L/
\sigma_{\mathrm{tot}}=0.076 \pm 0.013$. At next-to-leading order this yields $\alpha_S(36.6 
\ {\mathrm{GeV}}) = 0.150 \pm 0.025$.
The $\xi\equiv \ln (1/x)$ distributions were used  
in conjunction with OPAL data from 91-208 GeV to study the scale 
dependence of the maximum position, $\xi_0$. 
We studied flavour dependence of 
$\xi_0$ as a possible explanation of this problem,
using direct flavour dependent 
measurements of $\xi_0$ at 91 GeV by OPAL.

\end{abstract}

\maketitle

\section{$\sigma_L/\sigma_{\mathrm{tot}}$ with JADE data} 

\subsection{Introduction}

The energy and the momentum spectrum of a hadron $h$ produced in the annihilation 
process e$^+$e$^-$ $\rightarrow \gamma , \mathrm{Z}^0  \rightarrow h + X$ is described by a 
fragmentation function 
${\cal F}^h(x) \equiv (1/\sigma_{\mathrm{tot}})\cdot (\mathrm{d} \sigma^h/ \mathrm{d} x)$. 
With $x \equiv 2 p /\sqrt{s}$, $p$ is the momentum carried by the hadron $h$, 
and $\sqrt{s}$ is the centre-of-mass energy of the annihilation process with 
total hadronic cross-section $\sigma_{\mathrm{tot}}$. 
Ignoring polarisation effects 
the fragmentation function receives contributions from the 
transverse ($T$) and longitudinal ($L$) polarization states of the 
intermediate electroweak vector bosons, $\gamma$ and Z$^0$, and from their 
interference yielding an asymmetric contribution ($A$) \cite{bib-Nason-NPB421-473}.
% \begin{eqnarray}
%\label{eqn-costheta-dependence}
% \frac{1}{\sigma_{\mathrm{tot}}} \cdot
%              \frac{\mathrm{d}^2 \sigma^h}{\mathrm{d} x\ \mathrm{d}(\cos\theta)} =
%     \frac{3}{8}\left(1+\cos^2\theta\right)\cdot {\cal F}_T^h(x) \nonumber \\ 
%     + \frac{3}{4}\left(\sin^2\theta\right)  \cdot {\cal F}_L^h(x)
%     + \frac{3}{4}\left(\cos\theta\right)    \cdot {\cal F}_A^h(x)
%  \ \ \ ,
% \end{eqnarray}
% where $\theta$ is the polar angle between the direction of the incoming 
% and the outgoing hadron $h$. 
At centre-of-mass energies much larger 
than the mass of the produced quark $q$, the longitudinal contribution 
is negligible \cite{bib-BoehmHollik-CERN-89-08}. 
A sizeable contribution to the longitudinal 
fragmentation function comes from gluon radiation from the $q\bar{q}$ 
system in the final state \cite{bib-Nason-NPB421-473}. The asymmetric contribution
vanishes, because we don't distinguisch  
quark and antiquark. 
The fragmentation functions are related to the perturbatively calculable ratios 
of the longitudinal, $\sigma_L$, and transverse, $\sigma_T$, cross-sections 
to the total cross-section.
% Integrating Eq. (\ref{eqn-costheta-dependence}) over
%  $\cos\theta$ and respecting energy conservation for the integral over $x$ yields 
% \cite{bib-Nason-NPB421-473}
% \begin{equation}
% \label{eqn-sumrule}
%     \frac{1}{2} \sum_h\int \mathrm{d} x\  x\cdot 
%                \frac{1}{\sigma_{\mathrm{tot}}} \cdot
%                \frac{\mathrm{d} \sigma^h}{\mathrm{d} x} = 
%                         \frac{\sigma_{T}}{\sigma_{\mathrm{tot}}} + 
%                          \frac{\sigma_{L}}{\sigma_{\mathrm{tot}}}     = 1
% \ \ \ ,
% \end{equation}
% where
% \begin{equation}
% \label{eqn-sigma-def}
%      \frac{\sigma_{T,L}}{\sigma_{\mathrm{tot}}} \equiv 
%          \frac{1}{2} \sum_h\int \mathrm{d} x\  x\cdot {\cal F}_{T,L}^h(x)
% \ \ \ .
% \end{equation}
The contribution of gluon radiation to $\sigma_{T,L}/\sigma_{\mathrm{tot}}$
has been calculated in second order of $\alpha_S$ \cite{bib-Rijken-PLB386-422}. 
This allows tests of QCD and determinations of $\alpha_S$ from measurements
of $\sigma_L/\sigma_{\mathrm{tot}}$.

\subsection{Detector and data samples}

We present in this paper a re-analysis of data
recorded by the JADE detector \cite{bib-naroska}  at the PETRA electron-positron collider.
The data used for this study were recorded between 1979 and 1986 at 
centre-of-mass energies of $\sqrt{s} = 34$-$36$~GeV and $\sqrt{s} = 43$-$45$~GeV. 
Multihadronic events were selected according to the criteria described in
\cite{bib-Fernandez-EPJC1-461}.
For this analysis we used the JADE collaboration's original Monte Carlo samples 
of multihadronic events from the JETSET program version 6.3~\cite{bib-JETSET} 
including a detailed simulation of the JADE detector.

\subsection{Meaurement of $\mathbf{\cos\theta}$ distribution}
\begin{figure}
\includegraphics[bb=19 58 567 125,width=0.4\textwidth]{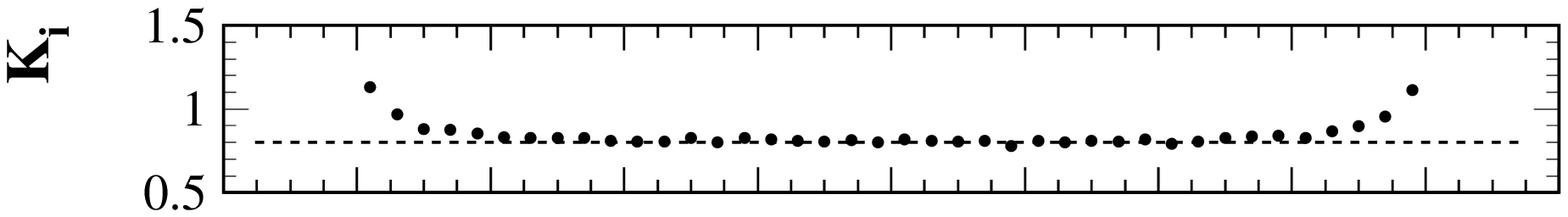} \\ 
\includegraphics[bb=19 10 567 510,width=0.4\textwidth]{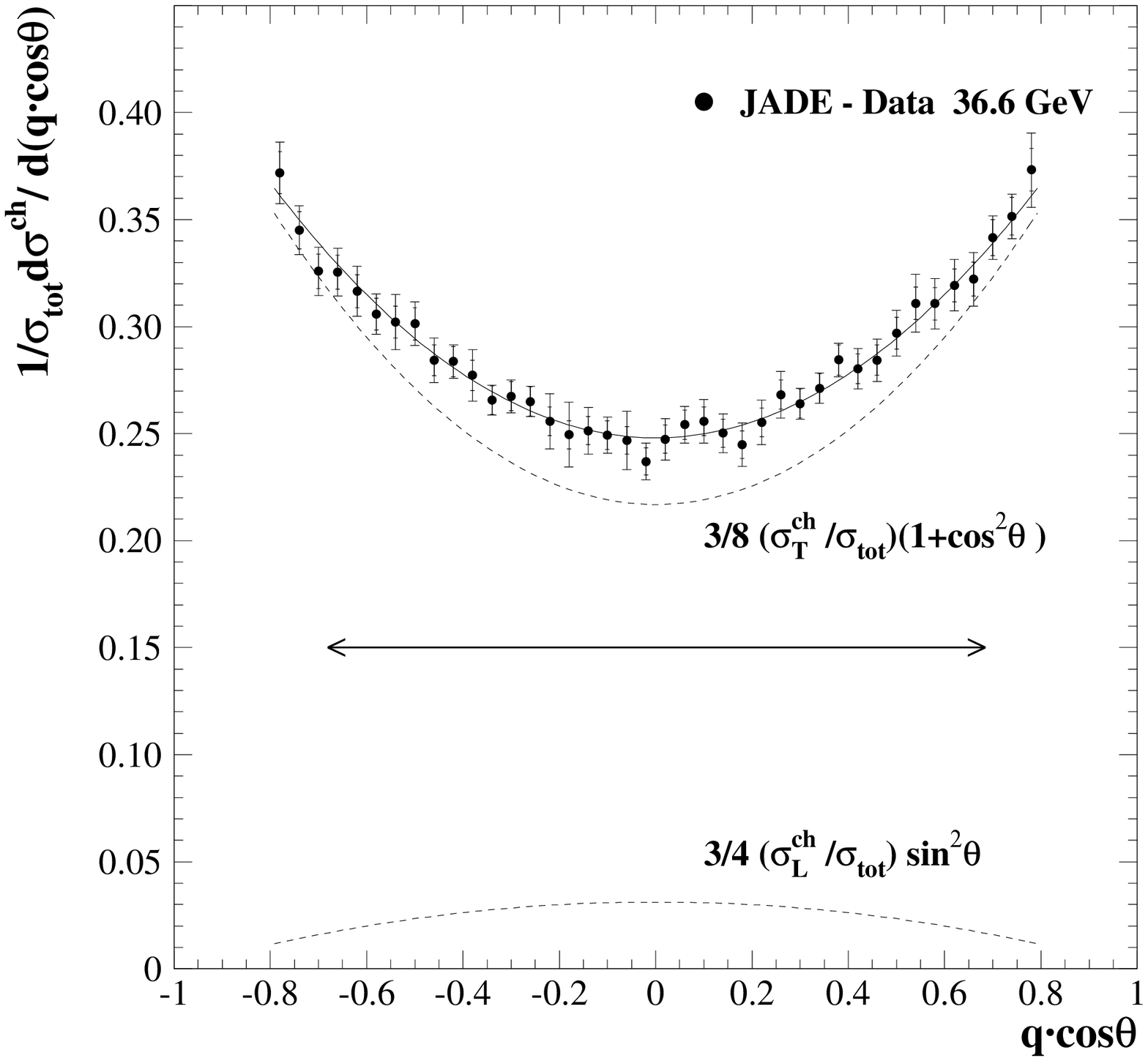} \\
\vspace{-1.cm}
\caption{\label{fig-costheta}
         The distribution of $q\cdot\cos\theta$ is shown after correction for 
         detector effects using the factors $K_i$ presented in the upper part 
         of the figure. The inner error bars are the statistical uncertainties 
	 and the outer bars are the total errors. The range considered for the 
  	 fit is indicated by the arrow.}
\end{figure}

The distribution 
of $\cos \theta$ of all tracks of charged particles was measured, where $\theta$ is 
the polar angle between the direction of beam axis and the outgoing hadron $h$.
The effects of limited acceptance and resolution of the detector were corrected
using a bin-by-bin correction method. 
Fig. \ref{fig-costheta} shows the measured distribution of $q\cdot\cos\theta$ 
after application of the binwise correction factors 
which are shown in the upper part of the figure.

\subsection{Determination of $\mathbf{\sigma_L/\sigma_{\mathrm{\bf tot}}}$
 and $\mathbf{\alpha_S}$}
\label{sec-observables}

Assuming that at the hadron level the correction for
neutral particles is identical for the longitudinal and transverse cross-sections, 
we can write :
% using Eq.~(\ref{eqn-sigma-def}) the differential cross-section for charged particles 
% given by Eq.~(\ref{eqn-costheta-dependence}) can be written as
\begin{eqnarray}
\label{eqn-sigmaL-fit}
\frac{1}{\sigma_{\mathrm{tot}}}\frac{\mathrm{d} \sigma^{\mathrm{ch}}}{\mathrm{d} (q\cdot\cos\theta)}
\ \ \ \ \ \ \ \ \ \ \ \ \ \ \ \ \ \ \ \ \ \ \ \ \ \ \ \ \ \ \ \ \ \ \ \ \ \ 
 \nonumber \\
= \frac{3}{8}\eta^{\mathrm{ch}}
               \left[\frac{\sigma_L}{\sigma_{\mathrm{tot}}}\left(1-3\cos^2\theta\right)
                                                         + \left(1+ \cos^2\theta\right)\right]. 
\end{eqnarray}
The unknown parameters to be determined from a fit to the data are 
$\eta^{\mathrm{ch}}$, the correction factor for the
total cross-section accounting for the neutral particles, and 
$\sigma_L/\sigma_{\mathrm{tot}}$. From 
                 substituting in Eq.~(\ref{eqn-sigmaL-fit}) with the relation known in 
${\cal O}(\alpha_S^2)$ \cite{bib-Rijken-PLB386-422}.
\begin{equation}
\label{eqn-QCD-prediction}
\left( \frac{\sigma_L}{\sigma_{\mathrm{tot}}}\right)_{\mathrm{PT}} =
                     \frac{\alpha_S}{\pi} + 8.444\left(\frac{\alpha_S}{\pi}\right)^2
\end{equation}
a formula can be obtained which 
allows for a direct determination of the strong coupling constant $\alpha_S$
at a renormalization scale $\mu=\sqrt{s}$.
Values of
\begin{eqnarray}
      \frac{\sigma_L}{\sigma_{\mathrm{tot}}} & = & 0.067 \pm 0.011 {\mathrm{(stat.)}} 
                                                      \pm 0.007 {\mathrm{(syst.)}}    \nonumber
\end{eqnarray}
were obtained
for the longitudinal and transverse cross-sections relative to the total hadronic 
cross-section.
Using the second order QCD prediction for the relative longitudinal cross-section,
a value of 
\begin{eqnarray}
      \alpha_S(36.6\ {\mathrm{GeV}}) & = &0.150 \pm 0.020 {\mathrm{(stat.)}}  \nonumber \\ 
                                             & \pm & 0.013 {\mathrm{(syst.)}} 
                                             \pm 0.008 {\mathrm{(scal.)}} 
\nonumber
\end{eqnarray}
was determined for the strong coupling.

\section{$\ln(1/x)$ distribution between $\sqrt{s}$ = 22 and 44 GeV}
\subsection{Introduction}
Even though the shape
of the momentum spectra cannot be calculated for the complete phase space, sound
predictions have been made for the shape and the energy evolution of the 
$\xi \equiv \ln(1/x)$ distribution \cite{bib-Dokshitzer-PLB115-242,bib-FongWebber-NPB355-54}.
Destructive interference for soft gluon emission suppresses the production of 
particles with very low momentum thus turning the $\xi$ distribution into an
approximate gaussian shape at asymptotic energies \cite{bib-Dokshitzer-PLB115-242}.  
The peak position, $\xi_{0}$, of this distribution is expected to 
depend in leading order linearly on $Y \equiv\ln(\sqrt{s}/2\Lambda_{\mathrm{eff}})$.
Here $\Lambda_{\mathrm{eff}}$ 
is related to the $\Lambda$ parameter of the running strong coupling constant 
but is not identical to it due to the approximations made in the calculation. 
Corrections in ${\cal O}(\alpha_S)$ to the asymptotic prediction yield
a skewed gaussian shape for the $\xi$ distribution next to its 
maximum \cite{bib-FongWebber-NPB355-54}:\footnotemark
\begin{eqnarray}
\label{eqn-FWeqn}
F_q(\xi,Y) = \frac{N(Y)}{\sigma\sqrt{2\pi}} \ \ \ \ \ \ \ \ \ \ \ \ \ \ \ \ \ \ \ \ \ \ \ \ \ \ \ \ \ \ \ \ \  \nonumber \\
\cdot \exp\left( \frac{k}{8} - \frac{s\delta}{2} - \frac{(2+k)\delta^2}{4}
                + \frac{s\delta^3}{6} + \frac{k\delta^4}{24}\right),
\end{eqnarray}
where $N(Y)$ is a normalization related to the multiplicity of charged
particles, $\delta \equiv (\xi - \langle\xi\rangle)/\sigma$ and
 \cite{bib-FongWebber-NPB355-54}
\begin{eqnarray}
   \label{eqn-ximean}
\langle\xi\rangle \equiv \langle\xi(Y)\rangle \ \ \ \ \ \ \ \ \ \ \ \ \ \ \ \ \ \ \ \ \ \ \ \ \ \ \ \ \ \ \ \ \ \ \ \ 
\nonumber \\
= \frac{Y}{2}\left(1+\frac{\rho}{24}\sqrt{\frac{48}{\beta Y}}\right)
                                    \cdot\left[1-\frac{\omega}{6Y}\right] + {\cal O}(1)
\end{eqnarray}
with $\beta\equiv 11-2N_f/3$, $\rho\equiv 11+2N_f/27$, $\omega=1+N_f/27$, $C_A=3$, $C_F=4/3$, and 
$N_f$ the number of active flavours, which is usually set to $3$ since gluons predominantly split 
into a pair of the lightest quarks (u, d, s).
The $y$-dependence of $\sigma$, $s$, $k$ and $N(Y)$ can be found in references 
\cite{bib-FongWebber-NPB355-54}.
The peak position $\xi_0(Y)$, is related to the
mean value by the relation\addtocounter{footnote}{-1}\footnotemark
\footnotetext{For simplicity the explicit $Y$ dependence of $\delta$, 
$\langle\xi\rangle$, $\xi_0$, $\sigma$, $s$, $k$, and $k_5$ is not exhibited.}
\begin{eqnarray}
\label{eqn-ximax}
        \xi_{0} - \langle\xi\rangle \approx \frac{3\rho}{32 C_A} \approx  0.35,
\end{eqnarray}
where the approximation is for large $Y$, and the numerical value is for ~$N_f=3$
\cite{bib-FongWebber-NPB355-54}.

\subsection{Detector and data samples}
\label{sec-detector}

For the measurement of the $\xi$ distribution, data recorded by JADE between 1979 and 1986 
at centre-of-mass energies of $\sqrt{s}=22$, $35$ and $44$~GeV are analyzed.  Our 
investigation follows the same lines as the measurement of the longitudinal and 
transverse cross-sections.
However we use here MC simulation data based on the PYTHIA 5.722 
event generator \cite{bib-PYTHIA} with parameters of \cite{bib-OPALtune}.
\subsection{Measurement of the $\xi$ distribudion }
\begin{figure}[]
\epsfig{file=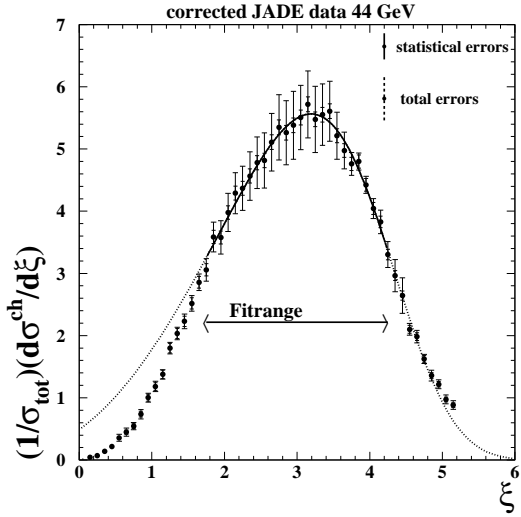,width=7.cm,clip=}
\vspace{-0.8cm}
\caption{\label{fig-44-lnxp}
         Measured $\xi$ distribution for 44 GeV corrected for detector
         effects. The inner error bars are the statistical uncertainties 
         and the outer bars are the total errors.}
\end{figure}
\begin{figure}[]
\epsfig{file=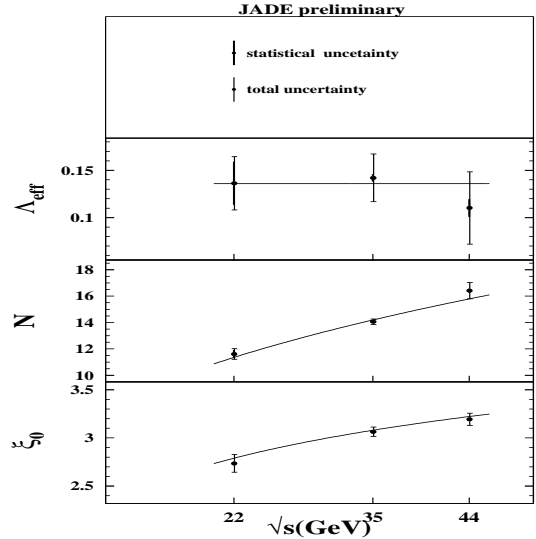,width=7.cm,height=7.5cm}
\vspace{-1.cm}
\caption{\label{fig-results}
         Results for $\Lambda_{\mathrm{eff}}$, $N$, $\xi_{0}$ versus the centre-of-mass
         energy $\sqrt{s}$. The curves  from the NLLA calculations are overlaid.}
\end{figure}
All charged particles  are used in the
measurement of the $\xi$ distribution. 
Fig.~\ref{fig-44-lnxp} shows the $\xi$ distribution measured from tthe 44 GeV data set
and corrected for the limited acceptance and resolution of the 
detector and for initial state radiation (ISR).
The position of the maximum $\xi_0$ is determined by fitting Eq.~(\ref{eqn-FWeqn})
to the data. 
The fit considers three parameters $N$, $\Lambda_{\mathrm{eff}}$ and $\xi_{0}$.
The remaining parameters of the skewed gaussian
(\ref{eqn-FWeqn}) are calculated using their $y$-dependance to be found in references
\cite{bib-FongWebber-NPB355-54}.
When fitting $\xi_{0}$ the 
asymptotic relation~(\ref{eqn-ximax}) is employed to substitute $\langle\xi(Y)\rangle$ in Eq.~(\ref{eqn-FWeqn}).
The final results of the fitted parameters \cite{bib-mona-note} including statistical and systematic uncertainties are summarized in Fig.~\ref{fig-results}. \\
Values of \\ 
$\xi_0(22 \mathrm{GeV})= 2.735 \pm 0.019 {\mathrm{(stat.)}} 
                                                      \pm 0.090 {\mathrm{(syst.)}}$\\
$\xi_0(35 \mathrm{GeV})=3.064 \pm 0.003 {\mathrm{(stat.)}}
 	\pm 0.049 {\mathrm{(syst.)}}$  \\
$\xi_0(44 \mathrm{GeV})=3.193 \pm 0.010 {\mathrm{(stat.)}}\pm 0.063 {\mathrm{(syst.)}}$\\
were obtained at PETRA energies of 22, 35 and 44 GeV.
\subsection{Investigation of flavour  dependence of $\xi$ distribution}
\begin{figure}[]
\epsfig{file=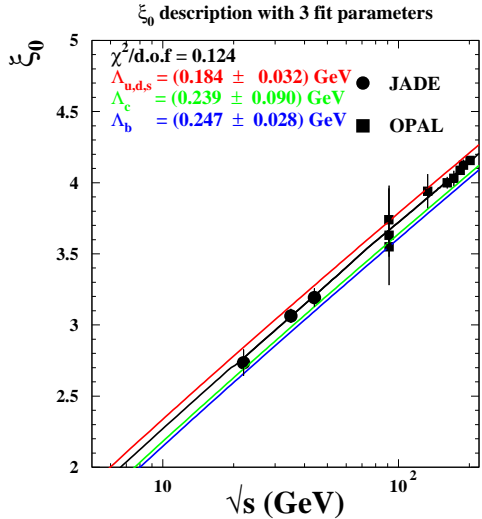,width=7.cm,clip=}
\vspace{-0.8cm}
\caption{\label{3lambdas}
         The measured position of the maximum $\xi_0$ is shown together
         with results at higher centre-of-mass energies. The curve is
          the expectation of QCD in NLLA, Eqs.~(\ref{eqn-ximax}) and (\ref{eqn-ximean}) with a flavour dependent $\Lambda_{eff}$, 
	this mean $\Lambda_{uds} \neq \Lambda_c \neq \Lambda_b$.}
\end{figure}

$\xi_0$ for the inclusive $\xi$ distribution can be written as a supperposition of $\xi_0$ for 
 flavour sorted $\xi$ distributions weighted with due to the different couplings 
of the intermediate photon and Z boson to the quarks center-of-mass energy dependent branching 
ratio for the corresponding flavour.
We consider the meassured values of $\xi_0$ from JADE data between $\sqrt{s}=$ 22 and 44 GeV
and OPAL data between $\sqrt{s}=$ 91 and 208 GeV. As constraints  we consider the directly 
meassured values of $\xi_0$ for $uds$, $c$ and $b$ at 91 GeV by OPAL \cite{Hapke}.
We assume that $\Lambda_{uds} \neq \Lambda_{c} \neq \Lambda_{b}$. 
With 3 fit parameters we described the energy evolution of 
$\xi_0$ for the inclusive $\xi$ distribution as shown in figure \ref{3lambdas}.
We meassured a differtent of 30$\%$ between 
$\Lambda_{b}$ and  $\Lambda_{uds}$.
\section{Summary and conclusions}
We measured with JADE data the longitutinal  
cross-section $\sigma_L/\sigma_{\mathrm{tot}}$ and the 
value of the strong copling constant $\alpha_S$ at 36.6 GeV and the peak position $\xi_0$ of the 
$\xi$ distribution at 22, 35 and 44 GeV.
Combining JADE and OPAL data we also investigated the flavour dependence of $\xi_0$.

\end{document}